\documentclass[letter]{aa} 
\usepackage{graphicx, comment}
\usepackage{commath}
\usepackage{txfonts}
\usepackage{subfig}
\usepackage{upgreek}
\usepackage[colorlinks=true]{hyperref}
\usepackage{soul}

\hypersetup{
     colorlinks   = true,
     citecolor    = blue
}

\usepackage{titlesec}
\titlespacing*{\section}{0pt}{1.2ex plus 0.8ex minus 0.2ex}{0.8ex plus 0.2ex}

\titlespacing*{\subsection}{0pt}{1.0ex plus 0.6ex minus 0.2ex}{0.6ex plus 0.2ex}

\titlespacing*{\subsubsection}{0pt}{0.8ex plus 0.4ex minus 0.1ex}{0.5ex plus 0.1ex}
\let\linenumbers\relax
\let\modulolinenumbers\relax
\let\runningpagewiselinenumbers\relax

\usepackage[normalem]{ulem}
\usepackage{orcidlink}
\newcommand{\Sp}{J1350$-$1634}
\newcommand{\Spg}{LEDA 896325}
\newcommand{\kms}{km~s$^{-1}$}
\newcommand{\msun}{$\rm M_{\odot}$}

\newcommand{\mjyb}{mJy~beam$^{-1}$}

\def\fs{\hbox{$.\!\!^{\rm s}$}}

\def\farcm{\hbox{$.\mkern-4mu^\prime$}}
\def\farcs{\hbox{$.\!\!^{\prime\prime}$}}
\begin{document} 
\title{Serendipitous discovery of a spiral host in a 2 Mpc double--double lobed radio galaxy}

\authorrunning{Sethi et al.}

\author{
Sagar Sethi\thanks{Corresponding authors;sagar.sethi@doctoral.uj.edu.pl;\\ agn.kuzmicz@uj.edu.pl}\fnmsep\inst{1,2}\orcidlink{0000-0001-8561-4228}
\and Agnieszka Ku\'zmicz\footnotemark[1]\inst{1,3} \orcidlink{0000-0002-3097-5605}
\and Dominika Hunik\inst{1} \orcidlink{0000-}
\and Marek Jamrozy\inst{1} \orcidlink{0000-0002-0870-7778}
}

\institute{
Astronomical Observatory, Jagiellonian University, ul. Orla 171, 30-244 Krakow, Poland
\and Doctoral School of Exact and Natural Sciences, Jagiellonian University, Krakow, Poland
\and Queen Jadwiga Astronomical Observatory in Rzepiennik Biskupi, 33-163 Rzepiennik Strzy\.zewski, Poland}

\date{Received -; accepted -}
\abstract
{
We present the serendipitous discovery of a double-double radio galaxy (DDRG) with a projected linear size exceeding 2 Mpc, hosted by a spiral galaxy. This unique combination of a giant radio structure and a spiral host challenges the prevailing view that such extreme radio sources reside only in elliptical galaxies. Using high-resolution optical imaging from the DESI Legacy Imaging Survey (DR10), we confirm a spiral-arm feature and a disk-component in the surface brightness profile fitting for the host galaxy (LEDA 896325) having a black hole of mass 2.4 $\times$ 10$^8$ \msun. Radio observations from RACS and GLEAM reveal two distinct pairs of radio lobes. Using the multi-frequency analysis of radio data, we obtained the spectral index distribution and estimate the spectral ages of the outer and inner radio lobes to be approximately 120 and 35 Myr, respectively. Our results confirm recurrent jet activity in this disk galaxy and establish it as the largest known radio galaxy in a spiral host, and its double-double structure makes it the largest of only three such spiral-host DDRGs, demonstrating that disk galaxies can indeed launch extremely large-scale radio jets.}

            
   \maketitle
\section{Introduction}\label{sec:intro}
Extragalactic radio jets, emanating from active galactic nuclei (AGNs) at the centres of some galaxies, are among the most powerful astrophysical phenomena \citep{AGN.Antonucci,Netzer_2013}. These are radio galaxies (RGs), often called radio-loud AGNs \citep[RLAGNs;][]{Urry.Padovani.RLAGN95}, which are predominantly bright in the radio band. Their jets can extend to large distances beyond the boundaries of host galaxies; those rare RGs that reach sizes larger than 0.7\,Mpc are called giant RGs \citep[GRGs; e.g.][]{GRG.Ishwara.Saikia,Kuzmicz18,Pratik.Rev23}. Jet activity typically persists for no more than a few tens of Myr and a RG in which jet activity has ceased is referred to as a `remnant' RG, where steep radio spectra can be observed in the outer lobes \citep{Jamrozy.Klein.2004, Parma.Murgia.2007, Murgia.Parma.2011, Brienza.2017, Quici.2021}. Some RGs also show evidence of episodic jet activity, with double-double lobed RGs \citep[DDRG;][]{Schoenmakers.2000} standing out as particularly striking examples of recurrent jet activity \citep[see reviews by][]{Saikia.Jamrozy.DDRG,Mahatma23.DDRG,Morganti24.DDRG}.

The vast majority of these RLAGNs are hosted predominantly by massive, gas‐poor elliptical galaxies that lack significant star formation \citep{Wilson.E.host,Urry.Padovani.RLAGN95}. In contrast, AGNs in spiral galaxies are typically radio-quiet with radio luminosities several orders of magnitude lower than those of RLAGNs. Spiral galaxies typically emit radio waves primarily through thermal free–free emission from the ionised interstellar medium (which rises at higher radio frequencies), along with synchrotron emission from their diffuse magneto-ionic medium \citep{Condon.1992}. The occurrence of RLAGNs with extended lobes in spiral galaxies is extremely rare. The first such object identified was a 200~kpc long radio source, J0313$-$192, whose host galaxy emits a 42~kpc long jet \citep{Ledlow.spiral.1998}.
So far, only a handful of unambiguous spiral host galaxies have been identified, including Speca \citep{Hota.spiral.2011}, J2345$-$0449 \citep{Bagchi.spiral.2014}, J0836$+$0532 \citep{Singh.spiral.2015}, J1649$+$2635 \citep{Mao.spiral.2015}, and MCG$+$07$-$47$-$10 \citep{Mulcahy.spiral.2016}. Moreover, only two of them --- Speca at z=0.137 and J2345$-$0449 at z=0.0755 --- extend to 1.4 Mpc and 1.6 Mpc, respectively, and each has multiple pairs of lobes indicating episodic AGN activity. It is worth mentioning that the sample of spiral-host RGs is being expanded recently \citep[e.g.][]{Wu.22,Yuan.24}. However, many of the selected objects require careful re-verification. Moreover, the spiral-hosted DDRGs and GRGs do stand out and they should be considered different from typical RGs and jetted Seyfert galaxies.

Analysis of the properties of known spiral galaxies with extended radio emission reveals several common features that characterize these unusual objects. One important property is the mass of the central supermassive black hole (SMBH), which in the case of spiral galaxies with powerful radio jets, is comparable to that typically found in elliptical galaxies \citep{Mulcahy.spiral.2016,Keel2006,Mao.spiral.2015}. For example, \citet{Bagchi.spiral.2014} found SMBH of up to 10$^9$~\msun \, and a rapidly rotating galactic disk, with velocities from 400 to 500 {\kms}. In the case of Speca, a rapidly rotating galactic disk with a velocity of 370 {\kms} was also found \citep{Hota.2014,Hota.2016}. The large mass of the black hole (BH) plays a key role in sustaining the activity of the central nucleus and producing the observed extended radio structures. Another important factor is the presence of interactions or mergers, which can trigger RLAGN activity in spiral galaxies, while preserving their spiral morphology \citep{Dorota.spiral.2012}. These interactions are thought to enhance the accretion of material onto the central SMBH, thereby powering the AGN and driving the production of radio jets. The surrounding environment also appears to be an important factor. Many of the RLAGN spiral galaxies are located in relatively dense environments, such as galaxy groups or clusters, which can provide the conditions necessary for extended radio emissions \citep{Singh.spiral.2015}. Interestingly, spiral galaxies with extended radio emission often have ongoing star formation and AGN activity. This coexistence suggests a complex interplay between the phenomena mentioned above \citep{Mao.spiral.2015}. Nevertheless, spiral RLAGNs, also known in the literature as Speca-like galaxies \citep{Hota.spiral.2011}, spiral-DRAGNs \citep{Mulcahy.spiral.2016} and/or late-type galaxies with double radio lobes \citep{Yuan.24}, remain exceptionally rare despite extensive efforts to find more of them. 

In this paper, we report the discovery of a spiral host {\Spg} in a GRG {\Sp} which extends over 2 Mpc in projected size. Furthermore, we have detected evidence for two distinct episodes of lobe activity using radio observations. As we show later, this RG surpasses all previously known spiral-host RGs in size. We briefly discuss the properties of the target and explore possible jet-launching mechanisms in this spiral galaxy.

Throughout the paper, we adopt the flat $\Lambda$CDM cosmological model based on the Planck Collaboration \citep[$\rm H_0 =67.8$ {\kms} Mpc$^{-1},\rm \Omega_m$= 0.308;][]{Planck16.Col}. We use the convention $\rm S_{\nu}\propto \nu^{\alpha}$, where S$_{\nu}$ is the flux density at frequency $\nu$ and $\alpha$ is a spectral index. The flux density scale is that of \citet{Baars.1977}. All positions and maps are given in the J2000.0 coordinate system. In the following, we will use the designation {\Sp} to describe the extended radio source, while we will refer to its optical host galaxy as {\Spg}.

\section{Target}
{\Sp} was initially identified as a candidate for GRG by \cite{Proctor} and later listed in the \citet{Kuzmicz18} GRG catalogue. It was classified as Fanaroff-Riley type II RG \citep[FR-II;][]{FR-type}, with an angular size of its radio structure of 11{\arcmin} measured on the NRAO VLA Sky Survey map \citep[NVSS:][]{nvss1}. {\Spg}, located at the position of RA: 13$^{\rm h}$50$^{\rm m}$36\fs10, Dec: $-16^{\circ}$34\arcmin50\farcs0, has been classified as either a quasar \citep{Kuzmicz18} or as a blazar \citep{wise_class}. It has a spectroscopic redshift of z$\rm _{spec}$=0.0877 \citep[6dF Galaxy Survey;][]{6dF}.

The region of the sky containing {\Sp} has extensive multiwavelength survey coverage (optical through radio), and we analyse the relevant data in more detail later (see in Sect.~\ref{sec:result_dis}). In addition, the galaxy is an X-ray source, appearing in the ROSAT Bright Source Catalogue \citep{Voges.1991} with a count rate of 8.58$\times10^{-2}$ s$^{-1}$ and its corresponding photon flux in the 0.1--2.4~keV band, when assuming a power-law energy distribution E$^{-1.3}$dE, is 3.13$\times10^{-12}$ mW~m$^{-2}$. This source is also clearly detected in the eROSITA all-sky X-ray survey \citep{Merloni.2024}. 

\begin{figure*}[h!]
\centering
    \includegraphics[scale=0.7]{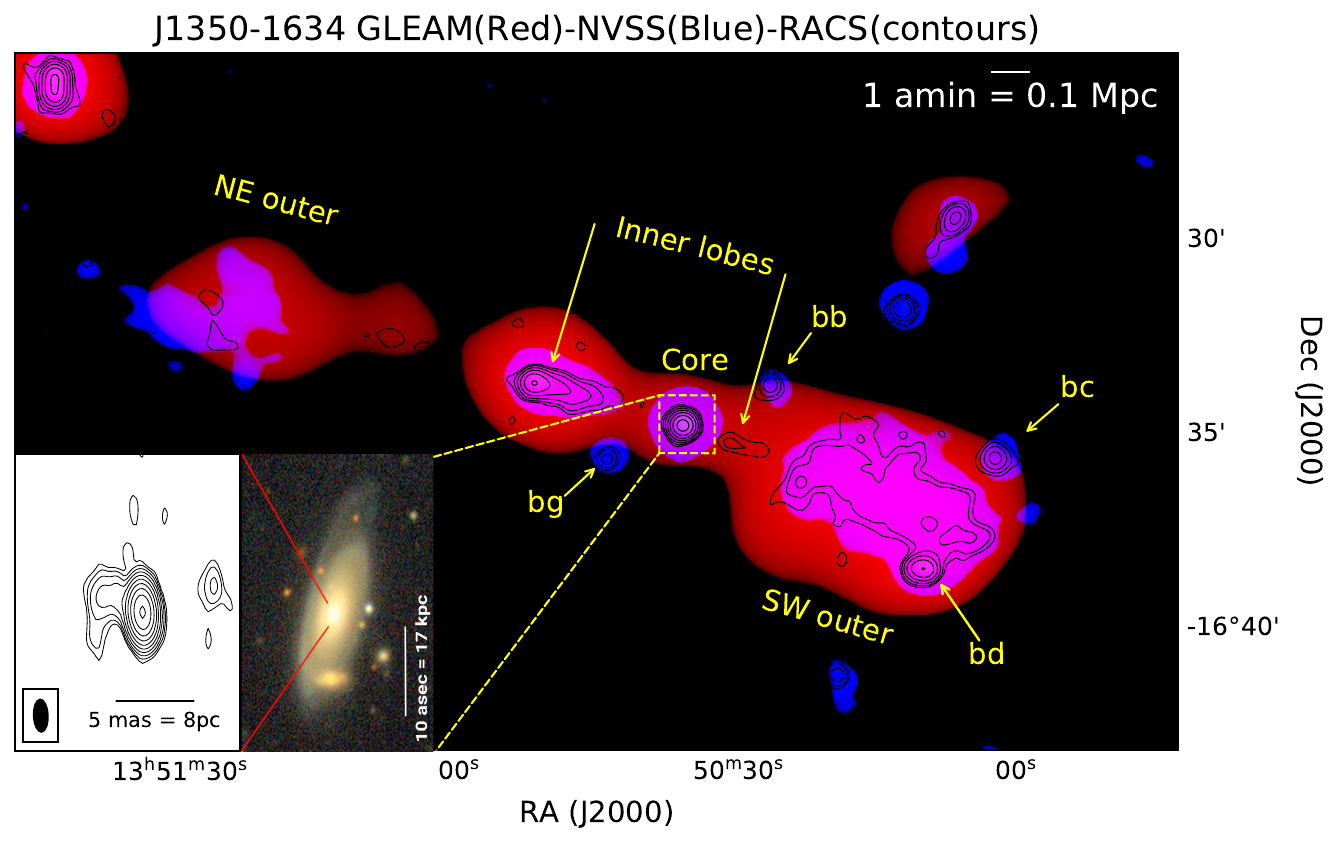}
    \caption{The false-colour image of {\Sp} is shown, where red and blue colours represent GLEAM (200\,MHz, with rms of 15\,\mjyb) and NVSS (1400\,MHz, with rms of 0.45\,\mjyb) respectively. The thin black contours of RACS-mid (887\,MHz, with rms of 0.3\,\mjyb) are plotted at $\rm 3\,rms\, \times\,2^n$ (n = 0,1,2,3 ...). Both red and blue pixels show values above 3 $\times$ rms for the respective data. The outer and inner lobes, as well as the core are marked with yellow. Background unrelated radio sources are labelled `bg', `bb', `bc', and `bd'. Two insets in the lower left corner show the colourful DESCal DR-10 optical image of the host galaxy {\Spg}, and the VLBI-scale radio core with one-sided jet, represented by 8.65 GHz contours.}
    \label{fig:radio}
\end{figure*}

\section{Results and discussion}\label{sec:result_dis}
To understand how a spiral galaxy can host such an extended RG, we have examined several aspects of J1350-1634. First, we examine the morphology of {\Spg} (Sect.\,\ref{subsec:host_morph}), then we estimate the central BH mass (Sect.\,\ref{subsec:bh_mass}). We consider possible triggers such as a companion or merger and the large-scale environment (Sect.\,\ref{subsec:merging}). Finally, we describe the radio structure that reveals the episodic jet activity (Sect.\,\ref{subsec:radio_morph}) and followed by spectral ages of both activity episodes (Sect.\,\ref{subsec:spectral_age}).

\subsection{The morphology of {\Spg}}\label{subsec:host_morph}
The high-resolution Dark Energy Spectroscopic Instrument (DESI) Legacy Imaging Surveys, Data Release-10 \citep[DESCal DR-10;][]{DESCal} image clearly reveals the spiral-like structure of the host galaxy, as shown in the lower left corner of Fig.~\ref{fig:radio}. The 6dF optical spectrum shows a relatively weak continuum and a broad H$\alpha$ emission line, suggesting that the host galaxy is more likely to be classified as a broad-line RG rather than a quasar.

\begin{figure}[]
\centering
    \includegraphics[width=\columnwidth]{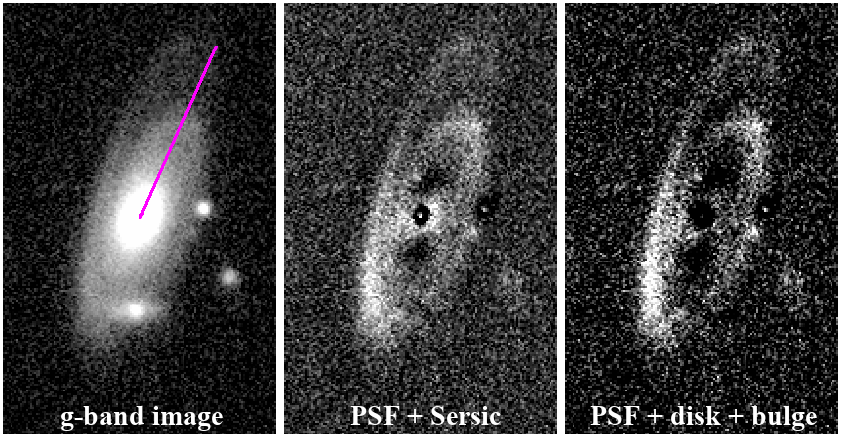}
    \caption{Left: The DESI g-band image of {\Spg}, with a magenta line indicating the cross section along the semimajor axis of the galaxy. Middle: The residual image for two components (PSF + Sersic from the whole galaxy). Right: The residual image for the three-component model (PSF + disk + bulge).}
    \label{fig:resid}
\end{figure}

\begin{figure}[]
\centering
    \includegraphics[width=\columnwidth]{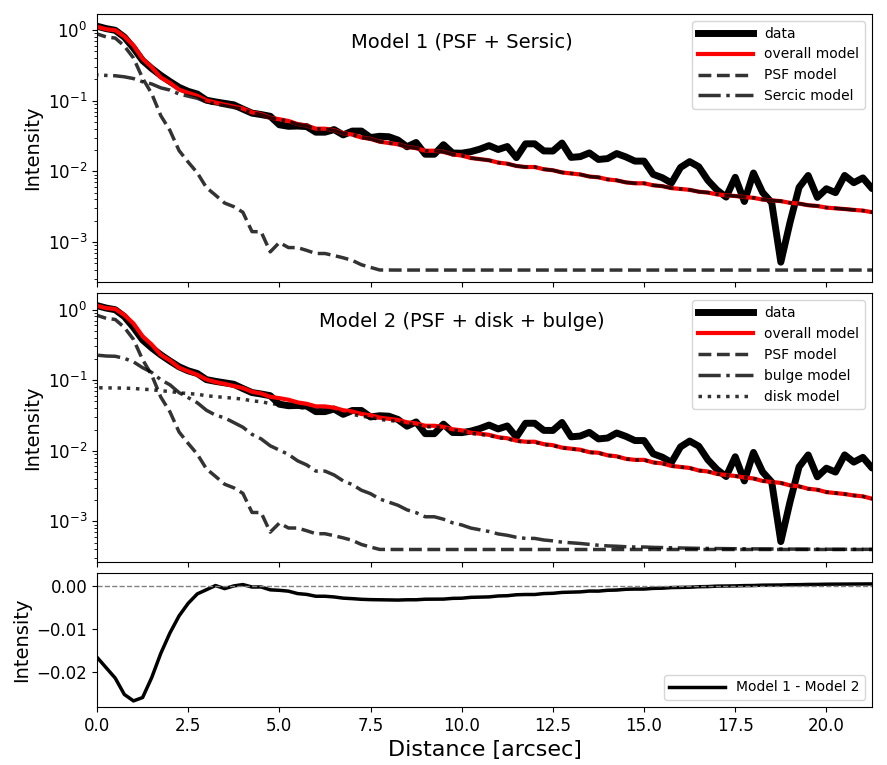}
    \caption{The surface brightness profiles along the cross-section, as indicated in Fig.~\ref{fig:resid}, for two different models: Model 1 (PSF + Sérsic) and Model 2 (PSF + disk + bulge). In each plot, we present the overall model together with its individual components. The bottom panel shows the residuals obtained by subtracting Model 2 from Model 1, highlighting the differences between these two approaches.}
    \label{fig:galfit}
\end{figure}

To analyse the morphology of LEDA 896325 and support its classification as a spiral galaxy, we modelled its surface brightness profile using a g-band DESCal DR-10 image. We performed a structural decomposition with GALFIT 3.0 \citep{Galfit}, fitting analytical models to the galaxy’s light distribution to separate its main structural components, such as the bulge and the disc, allowing for a quantitative characterisation of its morphology. We applied two different modelling approaches. In the first approach (Model 1), we used a two-component model with a point spread function (PSF) for the AGN contribution combined with a single Sérsic profile for the whole galaxy. The second approach (Model 2) uses a three-component model - a PSF for the AGN, an exponential disk, and a Sérsic profile for the bulge component. In both cases, the PSF for the AGN is included, since the optical spectrum of the host galaxy of {\Spg} has a broad H${\alpha}$ emission line, and the galaxy is a source of radio and X-ray emission. To remove the light contribution from foreground objects, our models includes two point sources visible in Fig.~\ref{fig:resid} (left panel), modelled using PSF profiles, and the foreground galaxy (labelled as `g2', see Fig.~\ref{opt_image}), overlapping {\Spg}, which was fitted with a two-component model consisting of a disk and a bulge. As a result of modelling, in the two-component model, we obtained the Sérsic index for the whole galaxy equal to n=1.8, indicating a morphology intermediate between a pure disk and a classical elliptical profile. In contrast, the three-component model yields a Sérsic index of n=0.8 for the disk component, consistent with an exponential light profile characteristic of galactic disks, and n=1.45 for the bulge, indicating the presence of a pseudo-bulge rather than a classical de Vaucouleurs bulge (n $\simeq$ 4). The residual images for both models, presented in Fig.~\ref{fig:resid} (middle and right panels), clearly reveal the spiral- or ring-like structure of the host galaxy. In Fig.~\ref{fig:galfit}, we show the surface brightness profiles along the galaxy's semi-major axis with both fitted models superimposed, including their individual components. Both modelling approaches give a comparable quality of fit to the data. To highlight the differences between them, we also show residuals obtained by subtracting one model from the other. The most significant discrepancies are observed in the region close to the central AGN. Although the two models provide different interpretations of the morphology of the galaxy, i.e. the two-component model suggests an intermediate structure between a spiral and an elliptical galaxy, while the three-component model classifies it as a disk galaxy with a pseudo-bulge, both models show that {\Spg} clearly exhibits disk-like features with spiral structures. This suggests that the galaxy has evolved by processes that have preserved its spiral structure, most likely through secular evolution or minor interactions, rather than major disruptive events.

\subsection{BH mass}\label{subsec:bh_mass}
To evaluate the BH mass of {\Spg} we derived the stellar velocity dispersion from the 6dF spectrum using the simple stellar population synthesis code STARLIGHT \citep{SSP.2005MNRAS}. This code models the observed spectrum by fitting the galaxy's spectral continuum with a combination of template spectra. As the 6dF spectra are not flux-calibrated, we normalized the spectrum of {\Spg} before modelling to minimize instrumental effects. Although STARLIGHT has been used previously to model 6dF spectra (e.g. \citeauthor{Gomes2012} \citeyear{Gomes2012}), its use is not recommended to infer stellar populations, ages, or galaxy masses due to the lack of  flux calibration. However, the code can be used for kinematic analyses, such as the determination of stellar velocity dispersion. After applying corrections for instrumental and stellar base dispersion, the resulting stellar velocity dispersion of {\Spg} is $\sigma_*$ = 240 \kms. Using the M$_{\rm BH}$ - $\sigma_*$ relation \citep{M.Sigma.Gebhardt} and adopting the constants from \cite{M.Sigma.Batiste.2017} derived for the AGN sample, we estimate the BH mass of {\Spg} to be M$_{\rm BH}$ = 2.4 $\times$ 10$^8$ \msun. This value is in the range of BH masses of giant elliptical galaxies, although spiral galaxies with central AGNs can also reach analogically high values. For example, \cite{Bagchi.spiral.2014} obtained a very similar value for the spiral galaxy J2345$-$0449, which is also a DDRG of size 1.6 Mpc. This supports the idea that a sufficiently massive BH is a crucial ingredient for the launch of extended radio jets even in a spiral host. While a high-mass BH is one key ingredient, external factors such as interactions could have also enabled RLAGN activity.

\subsection{Merging scenario and environment}\label{subsec:merging}
The possibility of interactions or mergers with neighbouring galaxies remains an important aspect of galaxy evolution. In the case of {{\Spg}}, the presence of an extended object located 12$^{\prime\prime}$ south of its centre (labelled as `g2' in Fig.~\ref{opt_image}) supports an interaction scenario. This object could represent a small companion galaxy that is currently undergoing an interaction or merger with the host galaxy. Such interactions could play a crucial role in shaping the dynamical evolution of the system, potentially facilitating bulge growth and triggering AGN activity \citep{Kormendy.Ho13}. However, the redshift of ‘g2’ remains unknown, leaving open the possibility that it could be a foreground object unrelated to the host.

Another interesting galaxy in the same region is LEDA\,896133 (labelled as ‘bg’ in Fig.~\ref{fig:radio} and \ref{opt_image}). This galaxy is located to the south-east of {\Spg} at an angular separation of 2\farcm13. Its photometric redshift is z$_{\rm phot}$ = 0.076 \citep{Bilicki.redshift}. Assuming that both galaxies are equidistant and that LEDA\,896133 has the same redshift as {\Spg}, their projected separation would be approximately 216\,kpc (i.e. about 3.5 times closer than the distance to M31). In particular, LEDA\,896133 has a lenticular / spiral morphology with some kind of ring-like structure and is also a bright and compact radio source (see Fig.~\ref{fig:radio}). These two large galaxies are most likely interacting.

In addition to these immediate companions, we examined the larger environment of {\Spg} to see if nearby galaxy groups or clusters might influence its AGN activity. We searched for galaxies using the GLADE+ galaxy catalogue \citep{GLADE+22} and for galaxy groups and clusters using the catalogues of \citet{Wen12}, \cite{Wen18} and \cite{Wen24} in the redshift range $z\pm\Delta z$ = 0.0877 $\pm$ 0.0033 (i.e. from 0.0844 to 0.0910). No galaxy cluster or galaxy group were identified within a radius of 6 Mpc (i.e. $1^{\circ}$) around the host, suggesting that {\Spg} resides in a relatively sparse environment. Only a few candidate galaxies with similar photometric redshift were randomly located around the host.  Future spectroscopic observations will be essential to confirm the merging scenario and to accurately characterize the surrounding environment.

\subsection{Radio morphology and episodic activity}\label{subsec:radio_morph}
The radio morphology of {\Sp}, as shown in Fig.~\ref{fig:radio}, reveals a complex structure that provides evidence of episodic jet activity. We noticed two pairs of lobes in {\Sp}, i.e. (1a) the north-eastern (NE) outer lobe, which was not recognized before and is more prominent in the Galactic and Extragalactic All-sky Murchison Widefield Array survey \citep[GLEAM;][]{GLEAM} but faintly visible in the NVSS; (1b) south-western (SW) outer lobe; (2a) SW inner lobe also before not recognized, visible only on the Rapid Australian Square Kilometre Array Pathfinder (ASKAP) Continuum Survey-low \citep[RACS--low;][]{racsL}; (2b) NE inner lobe, which is $\sim$2 times longer in angular size and $\sim$14 times brighter than its SW counterpart. These two newly recognized structures correspond to an outer and an inner pair of lobes, confirming that {\Sp} has undergone two distinct cycles of AGN activity. The largest angular size (LAS) of the outer lobes is 22\arcmin\,, corresponding to 2.24 Mpc, while the angular size of the inner lobes is 7\arcmin\,, corresponding to 0.71 Mpc. This makes {\Sp} not only the largest spiral host GRG, but also the largest spiral host DDRG known to date.

The well-separated radio core is clearly visible over a wide range of frequencies and its flux density increases with increasing frequency (see Table~\ref{table:flux_densities}). Its records for different epochs also suggest that it may be variable. The radio core is also visible in the Very Long Baseline Array (VLBA) map (see lower left panel of Fig.~\ref{fig:radio}), which was obtained from the Astrogeo VLBI FITS image database\footnote{\href{https://astrogeo.org/vlbi\_images/}{Astrogeo VLBI FITS image database}}. The VLBA core coincides exactly with the central position of {\Spg}. In the VLBA map there is also a NE extension which lies in the same direction as the NE outer lobe of this RG. This extension could be either a still visible jet from the latest activity or the beginning of  a new episode. If a new episode is confirmed, {\Sp} would be only the fifth known case of known ‘triple-double’ RG \citep{TDRG}. Further dedicated VLBA imaging or monitoring is required to confirm this. It may reveal a new episode or help us to monitor the jet's propagation. The counterjet is not visible at all, which may be due to the projection effect; the counterjet is away from our line of sight on the plane of the sky.

As mentioned above, the RG analyzed here is quite extended in the sky and there are several radio-loud objects in its vicinity. Most of them have low radio brightness that do not affect the flux density measurements of {\Sp}. However, several of them are noteworthy and should be taken into account. One of them is a background double RG (marked in Fig.~\ref{fig:radio} and in Table~\ref{table:flux_densities} as `bd') and two point-like radio objects (marked as `bb' and `bc' in Fig.~\ref{fig:radio}, while in  Table~\ref{table:flux_densities} only the brightest source `bc' is marked), where both the brightest objects are located at the head of the SW outer lobe (see Fig.~\ref{fig:radio}). {\Sp} has radio flux density measurements over a wide range of frequencies and the values are listed in Table~\ref{table:flux_densities}. These multi-frequency flux measurements allow us to derive the spectral index distribution and spectral ages of different lobes, as discussed next. These ages confirm two separate jet activities – consistent with the idea of recurrent activity in the same galaxy.

\subsection{Spectral Age}\label{subsec:spectral_age}
First, we obtained spectral index (SI) map of {\Sp} between 200 MHz (GLEAM) and 1400 MHz (NVSS). At the beginning, the highest resolution map was convolved to the resolution of the GLEAM map. Then the AIPS \citep{AIPS.Greisen} tasks \texttt{HGEOM} and \texttt{COMB} (with opcode `SPIX') were used to align the geometry of the two maps and finally to produce the spectral index map shown in Fig.~\ref{fig:radio_spectra} (left panel). In this map there are four well visible distinct structures, i.e. the core, the two outer lobes and the NE inner lobe. The core emission dominates the central part of the source, influencing and mimicking the SW inner lobe and the inner parts of the NE inner lobe. The core has a spectral index value of about 0. The SW outer lobe shows typical signs of an FRII-type spectral index behaviour, i.e. flatter spectrum with the value of about -0.8 at the front edge, where the presumed hotspot was located and smooth gradient of steepening spectral index towards the centre. This steepening towards the core is caused by the backflowing lobe plasma, where the aged plasma accumulates. The NE outer lobe also shows a steepening of the spectral index from the hotspot region towards the centre. The spectral index of the NE inner lobe is constant with the value of -0.8. However, there is a steeper region at its northern edges with a value of about -0.9 and this is probably related to the backflow of the NE outer lobe. The appearance of the spectral index described above is due to the radiation of relativistic charged particles in the 
presence of magnetic field. In the synchrotron process higher-energy particles lose energy much faster than their lower-energy counterparts. This causes the multiwavelength spectra to become increasingly curved over time.

We fitted the resulting observed radio spectra of the both outer lobes and the NE inner lobe (presented in Fig.~\ref{fig:radio_spectra}, right panel) using the SYNAGE package \citep{Murgia.1996}. Using the Jaffe \& Perola \citep[JP;][]{JP.1973} model, we estimated the radiative losses of the two outer lobes and the NE inner lobe, respectively. Assuming and fixing the injection spectral index value to 0.5, we evaluated the break frequencies to be 0.9 and 10 GHz for both the outer (SW \& NE) and the NE inner lobe, respectively. The fitting details were similar to those described by \cite{Sethi.2024}. To calculate the synchrotron age, we used, similar to e.g. \cite{Lusetti.2024}, the minimum magnetic field value of B$\rm_{min}=3.18(1+z)^2/\sqrt3$=2.2\,$\mu$G, which minimizes the radiative losses and maximizes the lifetime of the radio source, providing an upper limit. The resulting ages are 120 and 35 Myr for the outer and the NE inner lobes, respectively. These values of age, when compared with those for other DDRGs \citep[e.g.][]{Marecki.2021, Konar.2013}, appear as not special.
\section{Conclusion}\label{sec:conclusion}
In summary, we have identified {\Sp} as the largest of only three known spiral-host DDRGs with a size $>2$ Mpc. This exceptional source shows that spiral galaxies, although rarely, can launch and sustain powerful, large-scale radio jets without losing their disk morphology. The central BH of {\Spg} is as massive as those in giant ellipticals, likely facilitating its active jets, and the preserved spiral structure of the host suggests a gentle evolution (secular processes or small mergers rather than major disruptive events). These results challenge the traditional view that only ellipticals can produce giant RLANGs: given a massive BH and the right conditions, even a star-forming spiral can host episodic radio jets. This raises important questions: Why are such systems so uncommon? What special conditions or trigger mechanisms (e.g. particular interactions or environment) enable a spiral to ignite radio jets on the scale of millions of parsecs? Our discovery underscores the need for further investigation – both observational and theoretical – to understand how jets are launched in disk galaxies and how these galaxies maintain their structure. Continued further searches for similar objects and follow-up studies (e.g. deeper high-resolution radio imaging and spectroscopic observations of neighbours) will help to unravel the underlying processes that allow dual spiral+jet properties, shedding new light on AGN physics and galaxy evolution in unique environments.

\begin{acknowledgements}
{We thank the reviewer Prof. Ananda Hota for his very detailed and valuable comments that helped significantly improve the article. S.S., A.K., and M.J. were partly supported by the Polish National Science Centre (NCN) grant UMO-2018/29/B/ST9/01793. S.S. also acknowledges the Jagiellonian University grants: 2022-VMM U1U/272/NO/10, 2022-SDEM U1U/272/NO/15, and VMM-2023 U1U/272/NO/10. M.J. acknowledges access to the SYNAGE software kindly provided by Matteo Murgia. We gratefully acknowledge Polish high-performance computing infrastructure PLGrid (HPC Center: ACK Cyfronet AGH) for providing computer facilities and support within computational grant no. PLG/2024/016935 and PLG/2025/017961. We used in our work the Astrogeo VLBI FITS image database, DOI: 10.25966/kyy8-yp57, maintained by Leonid Petrov. We acknowledge \textsc{APLpy}\citep[][]{APLpy}, \textsc{astropy} \citep{astropy} and \textsc{matplotlib} \citep{matplotlib} being used in the paper to create all the plots.
}
\end{acknowledgements}

\bibliographystyle{aa}
\bibliography{ref}

\begin{appendix}
\onecolumn
\section{Optical image of {{\Spg}} galaxy}

\begin{figure}[ht!]
   \centering
    \includegraphics[width=0.8\columnwidth]{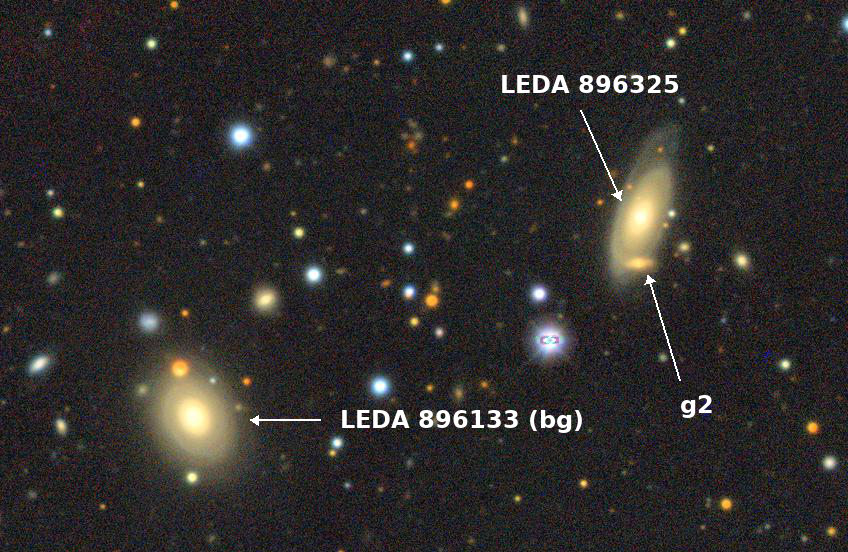}
    
      \caption{Optical image of the {\Spg} galaxy from the DESCal DR-10 survey, with potential companion galaxies labelled as `g2' and `bg'. }
         \label{opt_image}
\end{figure}

\section{Integrated flux densities of different components of {\Sp}}\label{sec:flux_table}
\begin{center}
\setlength{\tabcolsep}{3.5pt} 
\renewcommand{\arraystretch}{1.00} 
\scriptsize
\begin{table}
\captionsetup{width=\textwidth}
\caption{Observed flux densities of different components of {\Sp} and unrelated sources}
\centering
\begin{tabular}{rcccccccccc}
\hline
\tiny
Frequency       &\multicolumn{9}{c}{Flux density (mJy)} & Ref.\\
\multicolumn{1}{c}{(MHz)} & Total structure &  \multicolumn{3}{c}{$<$-------- Central structure ------$>$}& \multicolumn{2}{c}{Outer structure}&  \multicolumn{3}{c}{$<$------- Backg. sources$^{a}$ ------$>$}   \\  
&                                   & SW lobe     &   NE lobe      & Core       &  SW lobe & NE lobe      & bg   & bc   & bd  & \\
\multicolumn{1}{c}{(1)} & (2)                       &    (3)      &      (4)       &    (5)     &   (6)    &  (7)         & (8)  & (9) & (10)&(11)\\
\hline
        &               &               &             &                &            	 & 			 &          & &  &1  \\
84$^b$ 	&3294$\pm$335	&               &             &                &     2286$^c$$\pm$335& 575$\pm$176           &          & &  &1  \\
107$^b$	&2679$\pm$241	&               &             &                &     1746$^c$$\pm$241& 547$\pm$138           &          & &  &1  \\
130$^b$	&2118$\pm$184	&               &             &                &     1326$^c$$\pm$184& 441$\pm$105           &          & &  &1  \\
147.5	&               &               &155.3$\pm$13.1&29.6$\pm$3.2   &             	 &             	     	 &          &11.9$\pm$3.0&121.8$\pm$26.6&2   \\
158$^b$	&1645$\pm$147	&               &             &                &     1046$^c$$\pm$147& 277$\pm$82            &          & &  &1  \\
181$^b$	&1557$\pm$165	&               &             &                &      888$^c$$\pm$165& 365$\pm$104           &          & &  &1  \\
204$^b$	&1571$\pm$149	&               &             &                &      942$^c$$\pm$149& 341$\pm$93            &          & &  &1  \\	
224$^b$	&1379$\pm$172	&               &             &                &      826$^c$$\pm$140& 276.3$\pm$110          &          & &  &1  \\
888	&                       &4.97$\pm$0.86    & 71.5$\pm$3.9&83.2$\pm$4.2    &127.9$\pm$7.2&             		&4.99$\pm$0.42&13.9$\pm$1.0&42.5$\pm$2.3&3  \\
1368	&                       &               & 53.1$\pm$3.0&83.1$\pm$4.2    &            &             		&5.75$\pm$0.66& 8.09$\pm$0.92&29.5$\pm$1.5&4  \\
1400    &277.2$\pm$8.1          &               & 47.5$\pm$1.9&109.2$\pm$3.3   &101.5$\pm$6.5& 19.0$\pm$3.0            	&3.55$\pm$0.56& 5.53$\pm$0.60&29.5$\pm$1.3&5  \\
1656	&                       &               & 41.4$\pm$2.4&80.8$\pm$4.1    &            &             		&5.65$\pm$0.41& 6.67$\pm$0.64&23.0$\pm$1.2&6 \\
3000	&                       &               & 22.6$\pm$2.2&109.7$\pm$5.5   &            &             		&7.27$\pm$0.49& &20.4$\pm$1.9 &7  \\
4850    &208.3$\pm$28.3         &               &             &                &            &             		&          & &  &8  \\
5000	&                       &               &             &173$\pm$9       &            &             		&          & &  &9  \\
8000	&                       &               &             &220$\pm$12      &            &             		&          & &  &9  \\
8400	&                       &               &             &198$\pm$40      &            &             		&          & &  &10 \\
15000	&                       &               &             &173$\pm$15      &            &             		&          & &  &11   \\
20000	&                       &               &             &186$\pm$12      &            &             		&          & &  &9  \\
95060 	& 			& 		& 	      &188$\pm$20        &	    &                           &          & &  &12 \\
284990 	& 			& 		& 	      &79$\pm$8      &	    &                           &          & &  &12 \\
	&                       &               &             &                &            &             		&          & &  &   \\
\hline
\end{tabular}
\begin{flushleft}
\tablefoot{The columns are: (1) observed frequency; (2--10) flux density of the total structure of the target source (2), its inner lobes and the core (3--5), its outer lobes (6--7), and unrelated background radio sources (8-10); (11) references for the flux density measurements. {\em References} -- (1) GLEAM: \cite{GLEAM}; (2) \cite{tgss}; (3) RACS low: \cite{racsL}; (4) \cite{racsM}
(5) NVSS: \cite{nvss1}; (6) \cite{racsH}; (7) \cite{vlass1}; (8) \cite{pmn}; (9) \cite{AT20G}; 
(10) \cite{CRATES}; 
(11) average over 4 year of monitoring \cite{OVRO};
(12) \citet{ALMA.19.Bonato}. $^{a}$ the names of the background sources are marked in Fig.~\ref{fig:radio};
$^b$ the original flux density values from the GLEAM catalogue were averaged for three consecutive bands; $^c$ the flux density of the confusing sources `bc' and `bd' was subtracted from the original SW outer lobe flux density obtained from the catalogue.}
\end{flushleft}
\label{table:flux_densities}
\end{table}
\end{center}

\begin{figure*}[h!]
    \includegraphics[scale=0.45]{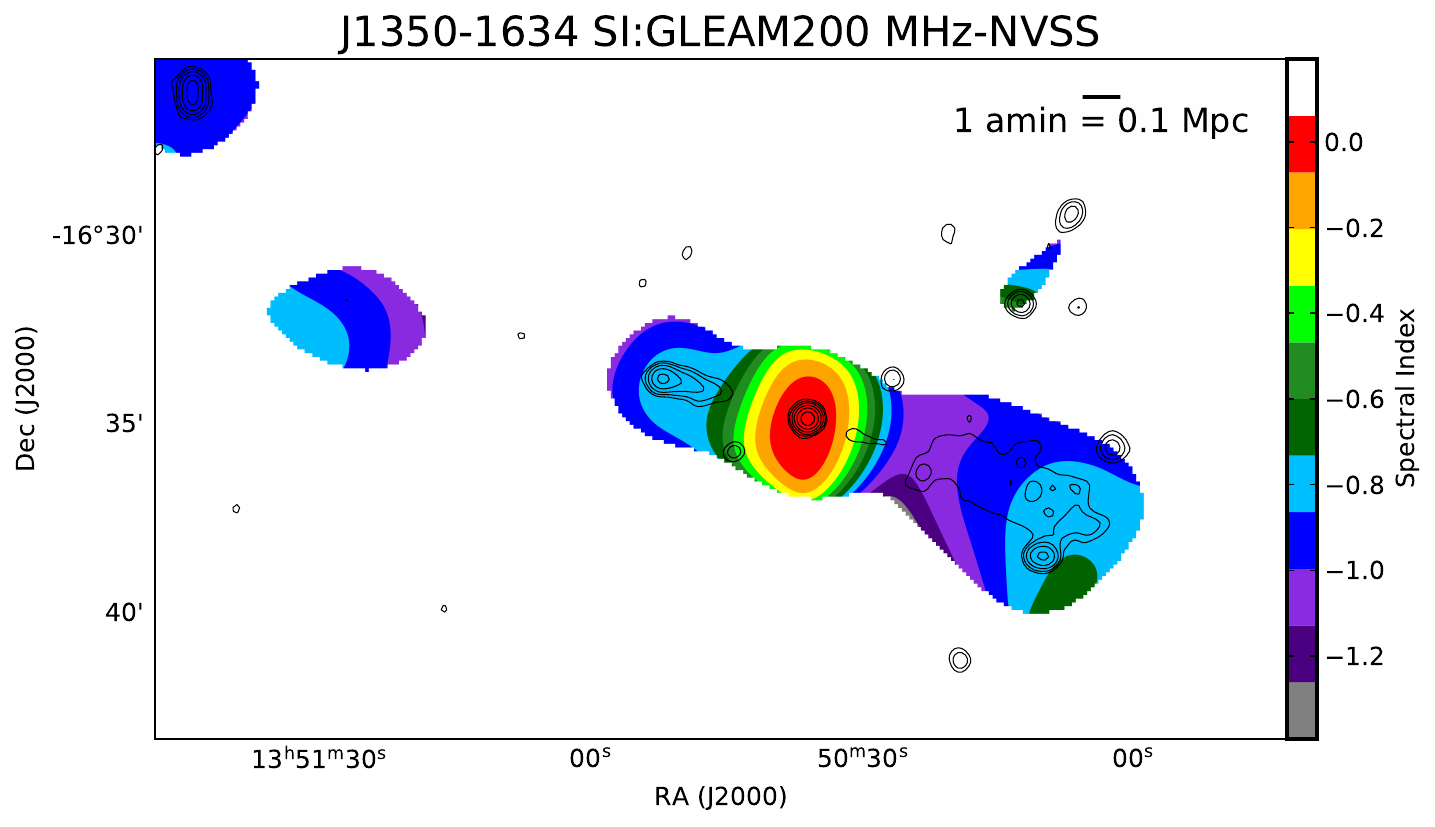}
    \includegraphics[scale=0.5]{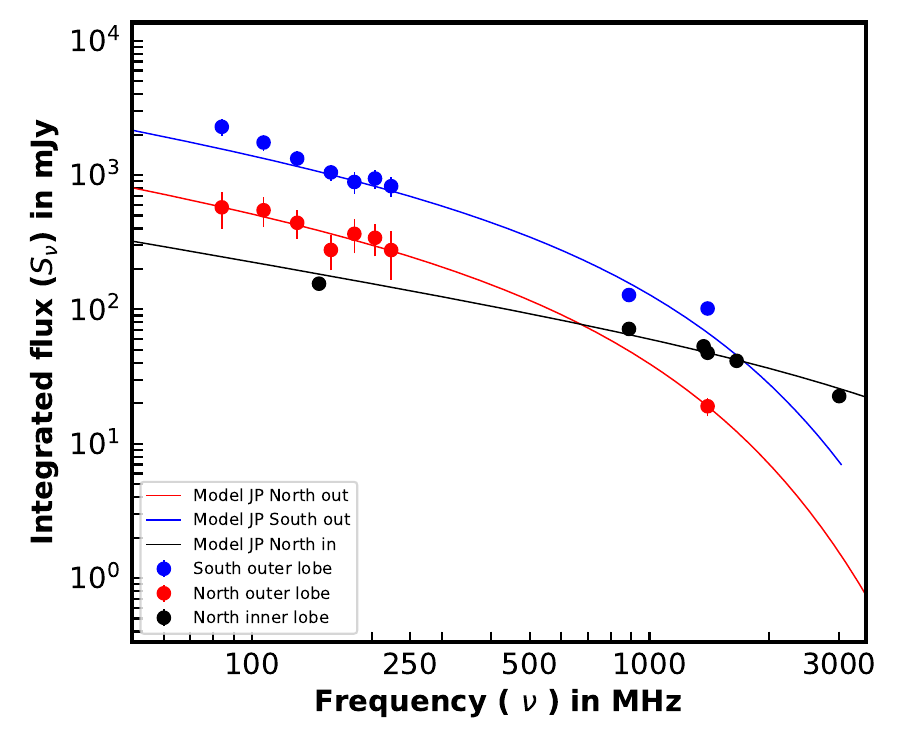}
    \caption{ The left panel shows the RACS-mid (887\,MHz, with rms of 0.3\,\mjyb) contours on the SI map between 200 MHz GLEAM and the convolved 1400 MHz NVSS maps. The  RACS-mid contours are plotted at $\rm 3\,rms\, \times\,2^n$ (n = 0, 1, 2, 3 ...). The right panel shows the radio spectra of both outer lobes (SW and NE) and NE inner of the target source fitted with the JP model with SYNAGE (for details, see Sect.~\ref{subsec:spectral_age}).}
    \label{fig:radio_spectra}
\end{figure*}
\end{appendix}

\end{document}